\documentclass[superscriptaddress,twocolumn]{revtex4}
\usepackage{amsmath,lscape,sidecap,epsfig}

\def\beq{\begin{equation}}
\def\eeq{\end{equation}}
\def\beqa{\begin{eqnarray}}
\def\eeqa{\end{eqnarray}}
\def\ban{\begin{eqnarray*}}
\def\ean{\end{eqnarray*}}
\def\bi{\begin{itemize}}
\def\ei{\end{itemize}}

\usepackage{graphicx,color,xcolor}

\begin{document}

\title{Pasta fluctuations in symmetric matter at finite temperature }

\author{Celso C. Barros Jr.}
\affiliation{Depto de F\'{\i}sica - CFM - Universidade Federal de Santa
Catarina  Florian\'opolis - SC - CP. 476 - CEP 88.040 - 900 - Brazil}
\author{D\'ebora P. Menezes}
\affiliation{Depto de F\'{\i}sica - CFM - Universidade Federal de Santa
Catarina  Florian\'opolis - SC - CP. 476 - CEP 88.040 - 900 - Brazil}
\author{Francesca Gulminelli}
\affiliation{CNRS and ENSICAEN, UMR6534, LPC, \\ 
14050 Caen c\'edex, France}

\begin{abstract}
The equilibrium distributions of the different pasta geometries and their linear sizes
are calculated from the mean field Gibbs energy functional in symmetric nuclear matter at finite temperature. The average sizes and shapes coincide approximately with the ones predicted by a standard pasta calculation in the coexisting phase approximation, but fluctuations are additionally calculated and seen to increase with temperature and baryonic density. 
The different pasta shapes are shown to coexist in a wide domain of density and temperature, in qualitative agreement with the findings of large scale molecular dynamics simulations, but with a much less expensive computational cost.
\end{abstract}
\maketitle

\section{Introduction}
Exotic non-spherical shapes of nuclear matter, the so called pasta
phases, are possible because of a competition between short range
nuclear attraction and long range Coulomb repulsion ~\cite{frustration}. 
Such complex phases are expected in the inner  crust of neutron stars (NS), as well as in core-collapse supernova cores~\cite{Ravenhall83,Hashimoto84,Horowitz04,Watanabe02,Maruyama05}. 

Even if their existence is limited to a very narrow range of densities, their electrical and thermal conductivity are believed to be very important for the thermal \cite{Brown09} and magnetic evolution of neutron stars~\cite{Pons13,Horowitz15}, and their elastic properties give some information on the quasi-periodic oscillations (QPOs)  observed in some soft-gamma ray repeaters (SGR)~\cite{Sotani11} and on magnetar giant flares~\cite{Horowitz10}.
Moreover, in supernova matter, the presence of density inhomogeneities contributes in an important way to neutrino opacity~\cite{pasta2bis,Horowitz15,Furtado16,Nandi18}. In particular, it was
suggested that pasta can slow neutrino diffusion in protoneutron stars and greatly increase the neutrino signal at late times after core collapse~\cite{Horowitz16}.

Most calculations of nuclear pasta assume a cristalline structure 
in the so-called Single Nucleus Approximation (SNA): different pasta geometries and sizes are associated to different temperatures, densities and proton fractions, but at a given thermodynamic condition matter is described by the spatial repetition of identical Wigner-Seitz cells. This is a poor approximation because at finite temperature different configurations can coexist at thermal equilibrium, and a rich phenomenology corresponding to multiple domains and defects is expected~\cite{Schneider16}. These fluctuations of composition are believed to be particularly important in the case of pasta, because of the very small energy barriers separating different geometries~\cite{Kubis19}. 
 In the calculations of crustal heating and cooling, these defects are modelled by an external impurity factor, which is essentially a free parameter~\cite{Daligaut09}.

The importance of fluctuations is confirmed by the realistic, and numerically very expensive, molecular dynamics simulations of nuclear pasta
: hundreds of thousands of particles are needed to avoid deformations due to boundary conditions and finite size effects~\cite{Dorso18,Horowitz18}. 
Self-consistent mean-field approximations with realistic effective interactions have also been performed~\cite{Newton09,Sebille,Okamoto13}, but no more than some thousand nucleons can be simulated even if massive parallel computing is employed~\cite{Okamoto13}. 
Shell and finite size effects were also investigated with the help of Skyrme-Hartree-Fock equations solved on a 3D Cartesian grid and some impressive structures were obtained. These effects were shown to be minimized only for very large number of nucleons, of at least $A=2000$ \cite{fattoyev2017}.  In these microscopic studies, it was observed that complex geometries different from the standard spheres, slabs and rods can be obtained, though at a very expensive computational cost. However, it is not clear if those triple periodic minimal surface (TPMS) pasta configurations \cite{schuetrumpf2} will survive at finite temperature, which is the object of the present study. For this reason, we stick to the standard pasta geometries in the following.
 
In a recent paper~\cite{Grams18},
a perturbative method was introduced, allowing to calculate the full nuclear distribution associated  to a given equation of state (EoS) of stellar matter based on the single-nucleus Wigner-Seitz approximation, with a numerical effort comparable to the one needed for a simple SNA. In this approach, the different configurations are weighted according to their Gibbs energy, which is consistently calculated from the EoS. The only hypothesis needed is that the corrections to the chemical potentials calculated in the SNA can be treated perturbatively, and neglected at first order. This hypothesis can be considered as safe in the high temperature domain investigated in the present work
, as it was recognized already in the eighties 
\cite{Burrows84}. This technique was applied in ref.\cite{Grams18} to evaluate the nuclear distribution during core collapse, based on 
a non-relativistic equation of state. In this paper, we use the same formalism but we apply it to a relativistic mean-field functional.   Furthermore, we use it for the first time to compute the distributions in the pasta regime, and show that at moderate temperature the different pasta geometries can coexist with comparable probabilities in a large range of densities, at very low computational cost. 

For this explorative study, we limit ourselves to symmetric matter. This has not a direct application for astrophysical purposes, but allows describing the main features of the pasta distributions as a function of density and temperature. The extension to asymmetric matter is in progress.

\section{Formalism}\label{formalism}
 
We consider a neutral system of electrons, positrons, protons and neutrons  
interacting with and through an isoscalar-scalar field $\phi$,  an isoscalar-vector field $V^{\mu}$ , and an isovector-vector field  $\mathbf b^{\mu}$  in a RMF approximation \cite{sw,pasta0}.

We use the NL3 parameter set, which is well known not to be one of the most realistic RMF models \cite{mariana_2014}. Nevertheless, in all previous works that use the same formalism \cite{pasta1,pasta2,pasta3,pasta2bis}, this model was chosen because of a reliable calculation of the surface tension coefficient, a very important quantity.
In future calculations, the parameterization will be reviewed and the proton fraction dependence will be also  investigated.
The equations of motion for the  fields are obtained and solved self-consistently  \cite{pasta0,pasta1,pasta2} 
and they depend on the  the  equilibrium  densities 
$\rho_B =\rho _{p}+\rho _{n} $,
$\rho_3 =\rho _{p}-\rho_{n} $, 
where $\rho _{p}$ and $\rho _{n}$ are the proton and neutron
densities, as well as on the associated scalar densities. The detailed expression of the different thermodynamic quantities can be found in refs.
\cite{pasta1,pasta2}.

As in \cite{pasta1,Maruyama05}, for a given total density $\rho_B$ and 
proton fraction $Y_p=\rho_p/\rho_B=\rho_e/\rho_B$ (with $\rho_e$ the net electron density),  
the average  characteristics of the pasta structures are built with different geometrical forms in a background nucleon gas. This is achieved within the so-called Coexisting Phase Approximations (CPA) by calculating 
from the Gibbs' conditions the equilibrium  
density $\rho^I$ and $\rho^{II}$ of the pasta and
of the background gas, as well as the 
proton fractions  $Y_p^I$ and $Y_p^{II}$ .

The total baryonic energy density of the mixed phase  is given by
\begin{equation}
{\cal E}_D= f {\cal E}^I + (1-f) {\cal E}^{II} + 
{\cal E}_{surf,D} + {\cal E}_{Coul,D},
\label{totener}
\end{equation}
where I and II label each of the phases, and $f$ is the volume fraction of 
the dense phase: 
\begin{equation}
f= \frac{\rho_B -\rho^{II}}{\rho^I-\rho^{II}}.
\end{equation}
In Eq.(\ref{totener}),  ${\cal E}^{I(II)}$ is the baryonic energy density of homogeneous matter at the baryonic density $\rho^{I(II)}$,
\begin{equation}
{\cal E}_D(\rho^{I(II)})={\cal E}_p^{I(II)}+{\cal E}_n^{I(II)}+{\cal E}_\sigma^{I(II)}+{\cal E}_\omega^{I(II)}+
{\cal E}_\rho^{I(II)} \label{ehomo}
\end{equation}
including both the terms associated to the fermions ${\cal E}_{n,p}$ and the ones associated to the fields ${\cal E}_{\omega,\sigma,\rho}$ \cite{pasta0}. This term only depends on the densities $\rho^{I(II)}$, $\rho_p^{I(II)}$, while
the surface and Coulomb term additionally  depend on the assumed geometry (D=1,2,3).
 By minimizing the sum  ${\cal E}_{surf,D} + {\cal E}_{Coul,D}$ with respect
to the size of the droplet/bubble, rod/tube or slab we get the virial condition
\cite{Maruyama05}
${\cal E}_{surf,D} = 2 {\cal E}_{Coul,D},$ and  
\begin{equation}
\label{eq:ecoul}
{\cal E}_{Coul,D}=\frac{2 \beta}{4^{2/3}}(e^2 \pi \Phi_D)^{1/3} 
\left({\sigma} D (\rho_p^I-\rho_p^{II})\right)^{2/3},
\end{equation}
where $\beta=f$ for droplets and $\beta=1-f$ for bubbles, 
 ${\sigma} $ is the surface energy coefficient,
$D$ is the dimension of the system. For droplets, rods and slabs,
\begin{equation} \label{eq:phid}
\Phi_D=\begin{cases}
\left(\frac{2-D f^{1-2/D}}{D-2}+f \right) \frac{1}{D+2}, \quad D=1,3;\\
 \frac{f-1-ln(f)}{D+2}, \quad D=2. \end{cases}
\end{equation}

We use the functional form given in \cite{pasta3} for the the surface tension coefficient, which was fitted from a full variation of the local density profile in the Thomas-Fermi approximation.

\begin{figure*}
\centering
\includegraphics[width=0.45\linewidth]{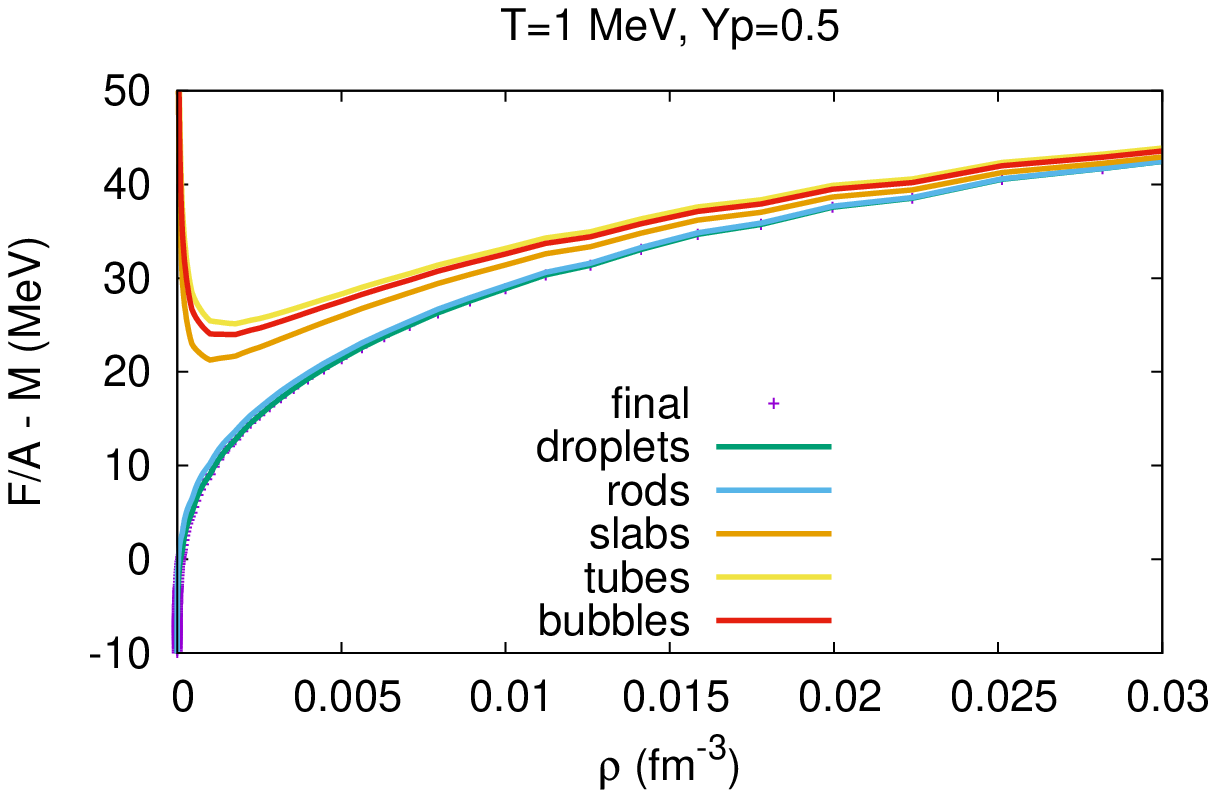}
\includegraphics[width=0.45\linewidth]{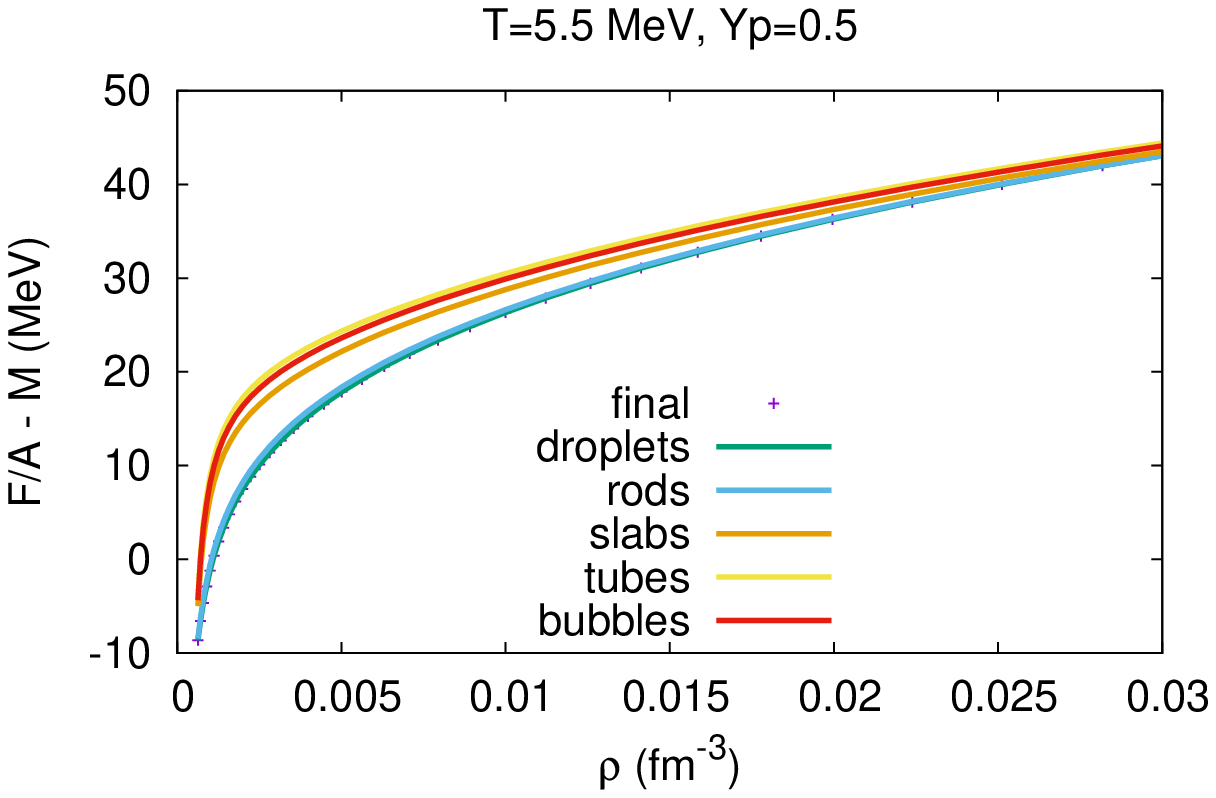}
\caption{Free energy per nucleon as a function of the density for $Y_p=0.5$ and $T=1$ and 5.5 MeV. }
\label{fig0}
\end{figure*}

Each structure is considered to be in the center of a charge neutral 
Wigner-Seitz  cell constituted by neutrons, protons and leptons (electrons and positrons) \cite{shen}. 
This average Wigner-Seitz cell
is a sphere/cilinder/slab whose volume is the same as the unit BCC cell. 
The linear size of the
droplet (rod, slab) and of the Wigner-Seitz cell are
respectively given by:
\begin{equation}
R_D=\left( \frac{\sigma D}{4 \pi e^2 (\rho_p^I-\rho_p^{II}
)^2 \Phi_D} \right)^{1/3}, \; \;
R_W=\frac{R_D}{f^{1/D}}. \label{eq:rd}
\end{equation}

The equilibrium pasta configuration is determined by comparing the total free energy per particle as obtained in the different geometries~\cite{Ravenhall83,Hashimoto84,Horowitz04,Watanabe02,Maruyama05,pasta0,pasta1,pasta2,pasta3}. As a general trend, the droplet shape prevails at the lowest densities, and is replaced successively by rods, slabs and bubbles as density increases. The location of the transition between the different geometries depends on the temperature, the proton fraction, and, to a smaller degree, on the nuclear model \cite{Providencia14}. 

In this formalism,  a single pasta configuration is associated to a given thermodynamic condition defined by ($T$, $Y_p$, $\rho_B$).
An example is given in Fig.\ref{fig0} top 
at two different temperatures, from where one can see the actual values of the free energy per particle in the different geometries and densities.
We can see that the free energy barrier between the different configurations is extremely low in a wide range of densities, and a qualitatively similar behavior is observed in the whole sub-saturation density domain. 
This is a generic result that was observed by many authors ~\cite{Ravenhall83,Hashimoto84,Watanabe02,Maruyama05} with different models. It means that we can expect a strong superposition of different shapes in a complete statistical calculation at finite temperature, as it is indeed observed in  large scale molecular dynamics simulations~\cite{Newton09,Sebille,Okamoto13,Schneider16,Horowitz18}. 
Indeed, in a complete statistical mechanics treatment of finite temperature matter, the total free energy should be minimized with respect to the 
probabilities of the different micro-states, and in the absence of long range correlations a statistical distribution of Wigner-Seitz cells is obtained~\cite{Gulminelli15,Grams18}. Working in the gran-canonical ensemble, the different cells share the same intensive parameters, namely the chemical potentials  
$\mu_n=\mu_n^{II}$, $\mu_p=\mu_p^{II}$ as well as the pressure $P=P^{II}$. As a consequence, the characteristics of phase II are not modified with respect to the CPA treatment.
Conversely, the density and proton fraction can fluctuate from one cell to another.

To avoid confusion with the definitions already used, 
the fluctuating density of a pasta structure is noted $\rho^N$ and the corresponding volume fraction $f^N$, with:
\begin{eqnarray}
\rho_B&=&\sum_N P^N f^N \left (\rho^{N} -\rho^{II}\right ) +\rho^{II} \\
\rho_p&=&\sum_N P^N f^N \left (\rho_p^{N} -\rho_p^{II}\right ) +\rho_p^{II} \  , \label{eq:rhoptot}
\end{eqnarray}
where $P^N$ is the probability of the cluster (high density) fluctuation corresponding to the density $\rho^N$, and $\rho^{II}$ refers to the more dilute phase of the pasta structure.

Charge conservation imposes $\rho_p=\rho_e$ for the global system, but charge neutrality is also realized at the level 
of the single Wigner-Seitz cell because of the homogeneity of the electron and proton gas,
  $\rho_e=\rho_{p}^N=Y_{p}^N\rho_{B}^N$.
For this first application, we 
neglect proton fraction fluctuations, i.e., we consider $Y_p^N=Y_p^{I}=Y_p^{II}=0.5$. Then, the baryonic density in each cell is the same, $\rho_B^{N}=\rho_B$, and we simply have:  $f^N=\frac{\rho_B -\rho^{II}}{\rho^N-\rho^{II}}$.

Since the SNA is known to reproduce very accurately the average thermodynamic properties, we make the approximation that the relation between chemical potential $\mu_n, \mu_p$ and total baryonic density $\rho_B$ can be obtained in the SNA, that is from the solution of the equilibrium Gibbs equations of the CPA.
The mean field equations 
require that each pasta configuration corresponding to a given density
$\rho^N$ leads to corresponding fluctuating meson fields.
 
The energy density is still given by eq.(\ref{ehomo}), but all the quantities are calculated at a density $\rho=\rho^N$.
For a given density value $\rho^N$, within the volume corresponding to the spherical cell $V_{W}^N =  4\pi   \left(R_{3}^N\right)^3/3f^N$, the number of particles of the cluster does not depend on the geometry and is given by
\begin{equation}
A^N={4\over 3}\pi \rho^N 
\left(R_3^N\right)^3  \  ,
\end{equation}
where $R_3^N$ is given by
\begin{equation}
R_3^N=\left( \frac{3\sigma }{4 \pi e^2 (\rho_p^N-\rho_p^{II})
^2 \Phi_3} \right)^{1/3}.
\end{equation}

Analogous expressions hold for the proton and neutron number $Z^N,N^N$ of the cluster.
For the other geometries, the same volume is fixed in our calculations, in which the number of particles is computed.

The probability of finding a pasta structure  with $A^N$ particles per unit cell, and dimension $D$ was demonstrated in ref.~\cite{Gulminelli15,Grams18} to be given by:
\begin{equation}
\label{eq:grams}
P_D (A^N)=\frac{\exp(-\beta \tilde G_D(A^N))}{\sum_D\sum_{A}\exp(-\beta \tilde G_D(A))}   \   ,
\end{equation}
such that the total probability of the fluctuation $\rho^{N}$can be
calculated as  $P^N=\sum_D P_D (A^N)$, where the effective one-body Gibbs potential is given by:
\begin{equation}
\label{eq:gnuc}
\tilde G_D(A^N)=\tilde F_D(A^N)-\mu_p Z^N-\mu_n N^N+ \delta F_D^N
\end{equation} 
and, in order to guarantee the statistical independence between the dense and dilute phase implied by eq.(\ref{eq:grams}), the Helmotz free energy of the cluster includes the interaction energy with the gas via the excluded volume term\cite{Grams18}:
\begin{eqnarray}
\tilde F_D(A^N)&=&\frac{{\cal E}_D(\rho^N)-{\cal E}(\rho^{II}) }{\rho^N} A^N  \nonumber \\
&-& T \frac{{\cal S}(\rho^N)-{\cal S}(\rho^{II})}{\rho^N} A^N ,
\end{eqnarray}
and ${\cal S}$ is the entropy density.
The rearrangement term $\delta F_D^N$ in eq.(\ref{eq:gnuc}) arises from the self-consistency induced by the Coulomb term \cite{Grams18}.
Indeed, the Coulomb term given by  eq.(\ref{eq:ecoul}) explicitly depends on the local proton density $\rho_p^N$ through the function $\Phi_D$ shown in eq.(\ref{eq:phid}). 
As the charge neutrality is realized at the level 
of the single Wigner-Seitz cell, its proton charge corresponding to the cluster $N$,
\begin{equation}
    \rho_p=f^N(\rho_p^N-\rho_p^{II})+\rho_p^{II},
\end{equation}
can be equalized to the global proton charge of the system given by eq.(\ref{eq:rhoptot}). To recover the same notations as in ref. \cite{Grams18} we can write eq.(\ref{eq:rhoptot}) in terms of the cluster densities $n_D(A^N)$ as:
\begin{equation}\label{eq:plocal}
    \rho_p=\sum_{N,D}n_D(A^N) V^N (\rho_p^N-\rho_p^{II}) +\rho_p^{II},
\end{equation}
where  $n_D(A^N)$ is linked to the cluster probability by:
\begin{equation}\label{eq:pglobal}
    n_D(A^N)=\frac{P_D(A^N)}{\sum_{N,D} P_D(A^N)V_W^N}=\frac{P_D(A^N)}{V_W},
\end{equation}
and $V_W$ is the average Wigner-Seitz cell volume, as given by the pasta calculation.
The equality between eq.(\ref{eq:plocal}) and eq.(\ref{eq:pglobal}) implies that the local proton density $\rho_p^N$
depends on the probability  of the associated fluctuation. As a consequence, a rearrangement term arises:
\begin{equation}
  \delta F_D^N=\left ( n_D \frac{\partial  \tilde F_D}{\partial n_D}\right ) (A^N). 
\end{equation}
Changing variables from $\rho_p^N$ to the volume fraction $f^N$ and using the virial relation 
${\cal E}_{surf,D}=2{\cal E}_{Coul,D}$ we have:
\begin{equation}
    \delta F_D^N=\frac{P_D(A^N)}{V_W}\frac{A^N V^N}{\rho^N}3\frac{\partial {\cal E}_{Coul,D}^N}{\partial f^N}.
\end{equation}

As $\tilde G_D$ should scale with $A^N$, 
we factor out the $A^N$ term, and average the rest of the expression over the different fluctuations, which amounts to replacing the different quantities with the ones obtained in the pasta calculation at the same thermodynamic conditions \cite{Grams18}:
\begin{equation}
    \delta F_D^N=A^N\frac{1}{\rho^I}3f\frac{\partial {\cal E}_{Coul,D}}{\partial f},
\end{equation}
which, after the replacement of eq.(\ref{eq:phid}), becomes:
\begin{equation}
\delta F_D^N=
\begin{cases}
\frac{A^N}{\rho^I} {\cal E}_{Coul,3} (3 + \frac{1}{5 \Phi_3}[f - f^{1/3}]),\quad {\rm droplets} \\    
\\
\frac{A^N}{\rho^I} {\cal E}_{Coul,2} (3 + \frac{(f-1)}{4 \Phi_2}), \quad {\rm rods} \\   
\\
\frac{A^N}{\rho^I} {\cal E}_{Coul,1} (3 + \frac{1}{3 \Phi_1} \frac{[f^2 -1]}{f}), \quad {\rm slabs}\\ 
    \end{cases}
\end{equation}

\begin{figure}
\centering
\includegraphics[width=1.0\linewidth]{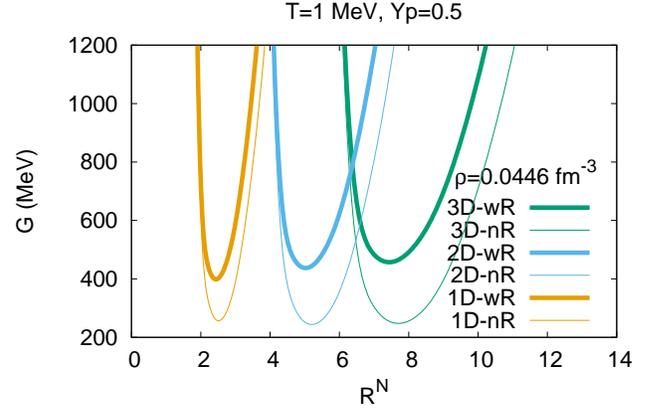}
\caption{Denser phase Gibbs free energy with (wR) and without (nR) the rearrangement term for T=1 MeV.}
\label{figrearr}
\end{figure}

 To quantify the rearrangement term, we plot the Gibbs energy with and without it for the three lowest geometries and two different densities at $T=1$ MeV in figures \ref{figrearr} bottom. We can see that  this term plays a non-negligible role and can even change the order of the preferential geometry.

\section{Results and discussion}\label{results}

To illustrate the formalism of section \ref{formalism}, 
we concentrate on the low density regime, close to the transition from the spherical to the rod shape, which is predicted by the NL3 model around $\rho_B=0.03$ fm$^{-3}$ at zero temperature and at slightly different densities as the temperature increases. 

\begin{figure}
\centering
\includegraphics[width=0.9\linewidth]{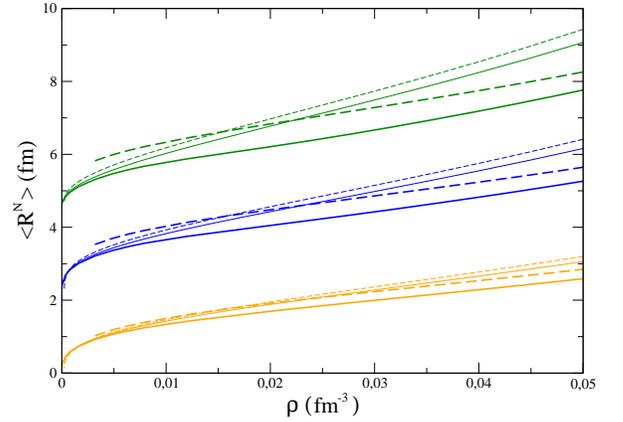}
\caption{Evolution with density of the linear size of the different geometries. 
Curves in green stand for 3D, curves in blue for 2D and curves in orange for 1D. Thick solid lines represent T=1 MeV, thick dashed lines T=4.5 MeV for average distribution radius; solid lines represent the pasta phase at 1 MeV and dashed lines the pasta phase for T=4.5 MeV.}
\label{fig4}
\end{figure}

Figure \ref{fig4} displays the evolution with density of the average linear size of the different geometries for two different temperatures. We can see that the temperature effects are negligible, while bigger pasta structures appear in denser matter, as expected. This is in perfect agreement with the standard pasta calculation in the SNA (thin solid and dashed lines in Figure \ref{fig4}), but in this latter a single geometry is considered in a given thermodynamical condition.

\begin{figure*}
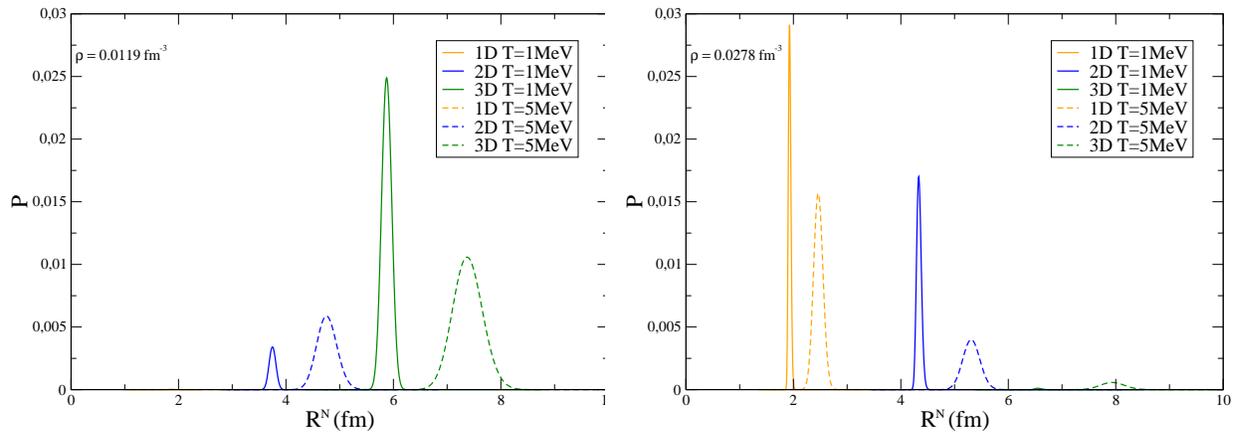

\centering
\includegraphics[width=0.45\linewidth]{fig4a_RC.eps}
\includegraphics[width=0.45\linewidth]{fig4b_RC.eps}
\caption{Probability distribution as a function of the pasta linear dimension with different geometries, for different temperatures and densities with
  $Y_p=0.5$.}
\label{fig1}
\end{figure*}

In our formalism, the different geometries can coexist and Figure \ref{fig1} shows the probability distribution as a function of the pasta linear dimension with different geometries and two temperatures, at two densities that correspond in the CPA approximation to the droplet and rod phase (see Figure \ref{fig0}). 
We can see that the droplet configuration (3D) dominates at the lower density and both temperatures considered, but the contribution of the rod geometry is far from being negligible. At slightly higher density, the situation changes: the slab configuration (1D) dominates at both temperatures and the three geometries can be seen to coexist at $T=5$ MeV. Whatever the geometry, the distribution is strongly peaked on the most probable cluster at the lowest temperature as expected, but considerable fluctuations are seen as the temperature increases.

\begin{figure}
\centering
\includegraphics[width=0.9\linewidth]{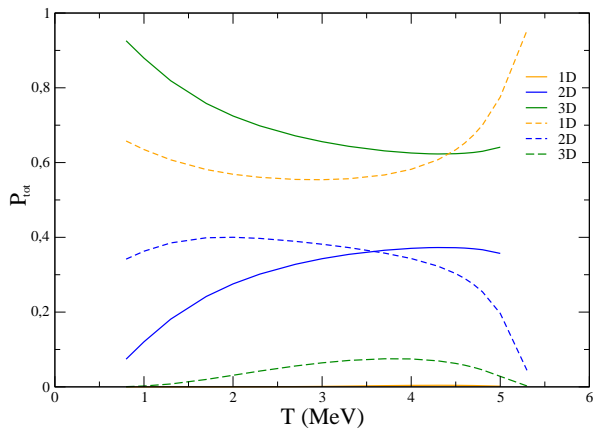}
\caption{Evolution with temperature of the probability of the
  different geometries. Solid lines represents $\rho$=0.0119
  $\rm{fm}^{-3}$ and dashed lines
represents $\rho$=0.0278 $\rm{fm}^{-3}$.}
\label{fig3}
\end{figure}

The evolution with temperature of the probability of the different geometries
is displayed in Fig. \ref{fig3} for two different densities.  
The droplet geometry tends to prevail at the low temperature in the lower density regime considered  
and remains dominant at all temperatures. At the lower density, the slab configuration (1D) is hardly noticed. At the higher density, the spherical configuration is replaced by the slab one as the dominant geometry and the other two geometries (3D and 2D) are also present, as already observed in figure \ref{fig1}. Hence, figure \ref{fig3} demonstrates that in a very wide range of temperatures the geometries coexist with comparable probabilities.

It is important to remark that in our formalism the pasta symmetry is always exactly respected. This is expected to be a good approximation at low temperature, well verified by microscopic calculations in three dimensions that allow all symmetry breakings \cite{Okamoto13}. 
However, more complicated geometries such as "waffle","parking garage" and "TPMS" structures \cite{Horowitz18,fattoyev2017,schuetrumpf2} are expected in some regimes of proton fractions, and could be added to the geometries presently considered.
This might  require extra corrections for applications in the supernova context where temperatures of the order of the MeV are explored. Indeed in this regime shape fluctuations are clearly observed in molecular dynamics simulations, even if finite thermalization time and finite size effects are difficult to handle and might distort the distributions \cite{Horowitz18,fattoyev2017}.

The application to the calculation of impurity factors for neutron
star cooling will be presented in a forthcoming paper. We would like to end by emphasizing the extremely low computational cost involved in our formalism as compared with \cite{Schneider16, Okamoto13, fattoyev2017,schuetrumpf2} where the coexisting geometrical shapes are observed. 

\section*{ACKNOWLEDGMENTS}
This work was  partially supported by Capes(Brazil)/Cofecub (France) 
joint international collaboration project number 853/15 and is part of the project INCT-FNA Proc. No. 464898/2014-5.
DPM acknowledges partial support also from CNPq (Brazil) under grant
301155/2017-8. DPM and CB thank the LPC-Caen for the hospitality.

\end{document}